\documentclass[%
 reprint,
showpacs,preprintnumbers,
 amsmath,amssymb,
 aps,
prb,
]{revtex4-1}
\usepackage[english]{babel}
\usepackage{graphicx}
\usepackage{dcolumn}
\usepackage{bm}
\usepackage[version=3]{mhchem}

\begin{document}

\title{{\textit{Ab Initio}} Study of the Early Stage of Si Epitaxy on the Chlorinated Si(100) Surface}

\author{Tatiana V. Pavlova$^{1,2}$}
\email{pavlova@kapella.gpi.ru}
\author{Egor S. Skorokhodov$^{1,3}$}
\author{Georgy M. Zhidomirov$^{1,2}$}
\author{Konstantin N. Eltsov$^{1,2}$}

\affiliation{$^{1}$Prokhorov General Physics Institute of the Russian Academy of Sciences, Moscow, Russia}
\affiliation{$^{2}$National Research University Higher School of Economics, Moscow, Russia}
\affiliation{$^{3}$Moscow Institute of Physics and Technology, Dolgoprudny, Russia}

\begin{abstract}
The homoepitaxial growth of Si on Si(100) covered by a resist mask is a
necessary technological step for the fabrication of donor-based
quantum devices with  scanning  tunneling  microscope lithography. In the present work, the chlorine monolayer is selected as the resist. Using density functional
theory, we investigated the adsorption of a single silicon atom on
Si(100)-2$\times$1-Cl as the starting process of Si epitaxy. The
incorporation of a silicon atom under a Cl monolayer proved to be the
most energetically favorable process. Our results show that chlorine
segregates on the surface during Si deposition and does not
incorporate into homoepitaxial layers. In addition, we found that
SiCl$_2^{\ast}$, SiCl$_3^{\ast}$, and SiCl$_4^{\ast}$ clusters can
be formed above a Si(100)-2$\times$1-Cl surface while Si is
adsorbed. SiCl$_2^{\ast}$ clusters are bound weakly to the
substrate, and their desorption leaves the silicon surface free of
chlorine. To check whether the Si epitaxy is possible on the chlorine resist, we compare our results with the well-studied case of a hydrogen resist. We find the two processes to be similar; moreover, epitaxy on chlorine resist appears to have an advantage.

\end{abstract}

\maketitle

\section{\label{sec:intro}Introduction}

Homoepitaxy of silicon is widely utilized for dopant encapsulation
in atomic scale devices. A sufficiently thick crystalline silicon
layer (about 50\,nm) should be grown on top of a surface with
dopants to preserve the electronic properties of semiconducting
devices from unwanted surface effects. Silicon epitaxy is a
necessary step for the fabrication of $\delta$-layers
\cite{2002Shen, 2015Keizer}, nanowires \cite{2012Weber, 2014Weber},
and quantum dots \cite{2012Fuechsle}. In addition, high quality of
epitaxial layers is extremely important for phosphorus-in-silicon
quantum computer building blocks \cite{2002Oberbeck, 2018Broome}.

To create the desired two-dimensional structure of impurities, a
silicon surface is covered with a resist which is then patterned (locally removed) with a scanning tunneling microscope (STM) tip. After the
adsorption of molecules containing specific impurities, impurity
atoms are embedded on the resist-free sites and then the surface is
covered with epitaxial silicon layers. The quality of epitaxial
layers strongly depends on the interaction of Si adatoms with atoms
of the resist used for mask fabrication. If a hydrogen monolayer is
used as a mask for the patterning of  Si(100)-2$\times$1 surface,
epitaxial layers of good quality are obtained by silicon overgrowth
at 250$^{\circ}$C \cite{2002Oberbeck,2016Deng,2018Hagmann}, but
dopant diffusion is not negligible \cite{2018Hagmann}. However, to control the positions of the dopants precisely, it is necessary to prevent the movement of the P donors during silicon overgrowth. Although dopant diffusion can be minimized by growing several locking layers at room temperature prior to higher-temperature Si epitaxy \cite{2015Keizer,2018Wang,2018Hagmann}, this partially suppresses the electrical activation of the dopants \cite{2015Keizer,2018Hagmann}. There is a challenge in looking for an optimal solution, retaining the dopants in their original embedded places without suppressing their activation.

Theoretical calculations of Si atom adsorption and diffusion on
H-terminated Si(100)-2$\times$1 surface have provided an insight into the
mechanism of homoepitaxy \cite{1997Nara, 1997Jeong, 1998Jeong}. A
silicon atom adsorbed on Si(100)-2$\times$1-H can spontaneously
substitute a hydrogen atom on the surface and then forms Si
dihydride. Experimental results confirm the presence of surface Si
dihydrides at submonolayer coverages \cite{2005Kajiyama}. As such a
Si film grows, most H atoms segregate on the surface and are not
incorporated into the epitaxial film \cite{2004Ji, 2016Deng}.

A chlorine monolayer on a silicon surface can also be utilized as a
resist \cite{2007Moon, 2011Jeon}. The key difference between a
hydrogen and chlorine resist is the potential possibility to remove
substrate atoms (Si) together with resist (Cl) atoms by the STM tip
because of the strong interaction of chlorine with silicon. This claim
is supported by a well-known etching effect that chlorine demonstrates
on Si(100), studied both theoretically \cite{1998deWijs} and
experimentally \cite{2009Aldao}. Of particular interest is a
proposal to use STM lithography on Si(100)-2$\times$1-Cl for placing
P atoms with atomic precision \cite{2018Pavlova}. After positioning
a P atom on a patterned Si surface, we have to build the silicon
lattice with an additional Si layer of 30--50\,nm as for
Si(100)-2$\times$1-H case. Si homoepitaxy directly on a chlorinated
Si surface looks a reasonable way to do it. While the interaction of chlorosilanes and HCl with silicon surface was investigated to find the mechanism of silicon epitaxy \cite{2001Hall, 2018Kunioshi, 2019Yadav}, and the interaction of various molecules with chlorinated silicon surfaces was considered for the development of functionalized surfaces \cite{2014Gao,2016Gao} and for the investigation of the reactivity of chlorinated silicon surface \cite{2008Lange,2013Soria}, there are no theoretical or experimental studies of the adsorption of silicon atoms on a Cl-terminated Si(100)-2$\times$1 surface.

In this paper, we report the results of density functional theory
(DFT) calculations of Si atom adsorption on Si(100)-2$\times$1-Cl. A
silicon adatom spontaneously substitutes a Cl atom and further
migrates to the most stable inter-bridge dimer site (bound with two
Si and two Cl atoms). This process is similar to that in the case of
Si(100)-2$\times$1-H surface, and should lead to Cl atoms
segregation during Si epitaxy. Despite the small radius of hydrogen
being suggested \cite{1997Jeong} as a reason for spontaneous
substitutional adsorption of silicon on Si(100)-2$\times$1-H, the
same adsorption mechanism is found to work for a chlorinated silicon
surface. Moreover, there is an additional pathway of spontaneous Si
adsorption with formation of SiCl$_2^{\ast}$ clusters weakly bound
with the silicon surface. Desorption of SiCl$_2^{\ast}$ clusters
from the surface requires low activation energy, so chlorine can be
removed without annealing.

\section{Calculation method}
First-principle calculations of silicon atom adsorption on
Si(100)-2$\times$1-Cl surface were performed with spin-polarized
density functional theory implemented in VASP \cite{1993Kresse,
1996Kresse}. The generalized gradient approximation with the
exchange-correlation functional in the form of
Perdew--Burke--Ernzerhof (PBE) was applied  \cite{1996Perdew}. The
eigenfunctions of valence electrons were expanded in a plane waves
basis set with an energy cutoff of 350\,eV. The Si(100) surface was
simulated by recurring 4$\times$4 cells, each consisting of eight
atomic layers of silicon. The slabs were separated by vacuum gaps of
approximately 15\,{\AA}. The bottom three layers were fixed at bulk
positions, while the other silicon layers were allowed to relax. The
lowest layer was covered by hydrogen atoms to saturate the dangling
bonds of silicon. Chlorine atoms were placed on the upper side of
the slab to form a Si(100)-2$\times$1-Cl structure. Reciprocal cell
integrations were performed using the 4$\times$4$\times$1 k-points
grid.

The adsorption energy ($E_{ads}$) of a silicon adatom was calculated
as the difference between the total energy of the surface with the
adatom ($E_{Si+surf}$) and the total energies of the
Si(100)-2$\times$1-Cl surface ($E_{surf}$) and a Si atom in the
gaseous phase ($E_{Si}$):
\begin{equation}
E_{ads} = E_{Si+surf} - E_{surf} - E_{Si}. \label{eq:1}
\end{equation}

The activation barriers ($E_{act}$) were calculated using the
climbing nudged-elastic band (CI-NEB) method \cite{2000CNEB} with
six images (including the two end points).

\section{Results}

\begin{figure}[h]
\begin{center}
    \includegraphics[width=\linewidth]{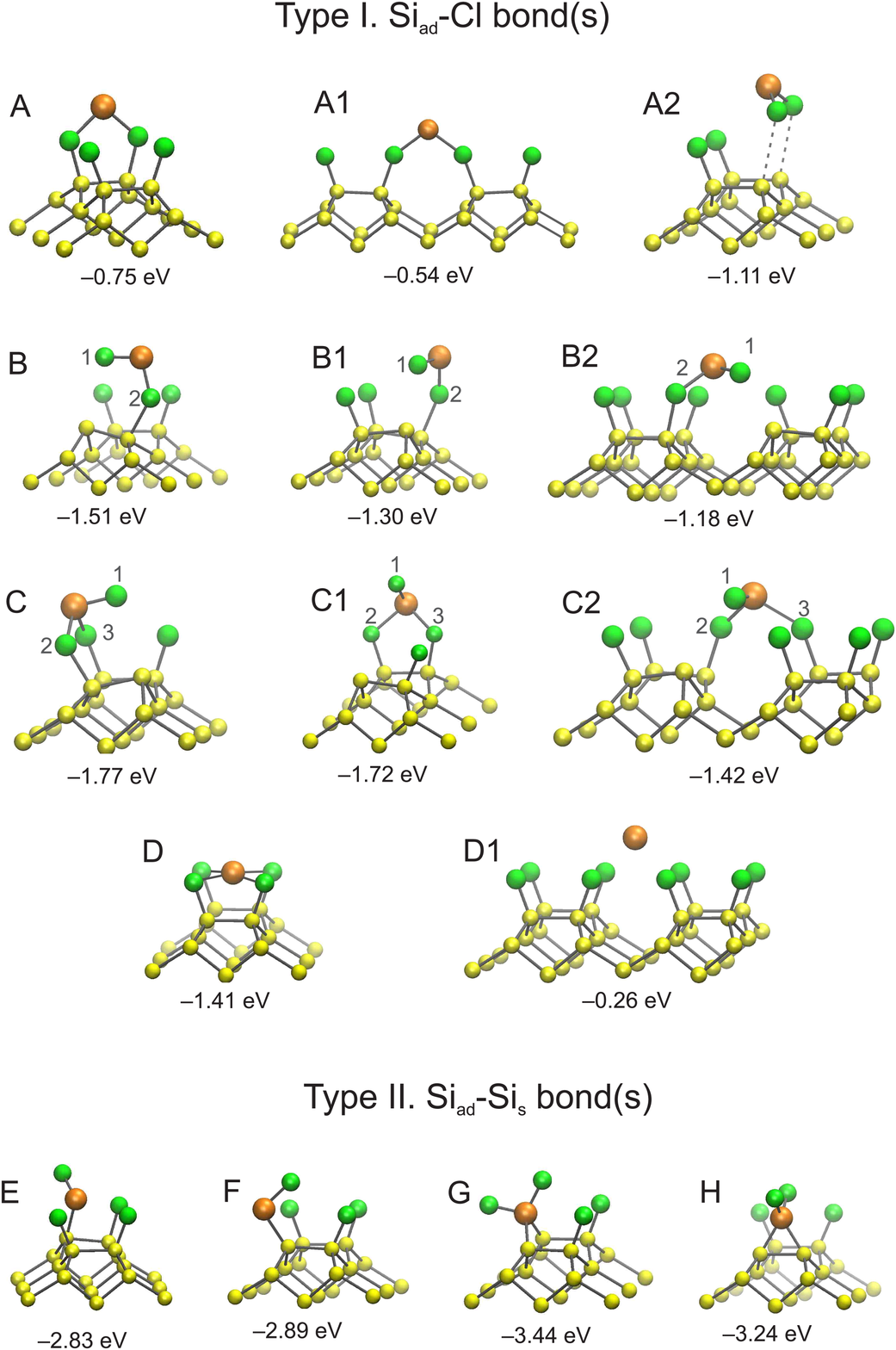}
    \caption{Optimized atomic structures for Si$_{ad}$ adsorbed on Si(100)-2$\times$1-Cl. Adsorption energies are indicated for every structure. Silicon atoms are yellow, chlorine atoms are green, and Si adatom is orange.}
    \label{fig1}
\end{center}
\end{figure}

Figure~\ref{fig1} shows the obtained adsorption structures for Si adatom
(Si$_{ad}$) on a Si(100)-2$\times$1-Cl surface. There are two types
of adsorption structures: Si$_{ad}$ bound with Cl atoms (type I) and
Si$_{ad}$ bound with Si surface atom(s) (Si$_{s}$) (type II). The
structures of type II are more energetically favorable, therefore
the formation of Si$_{ad}$--Si$_{s}$ bonds stabilizes the surface structure. We discuss further only one structure of type I for each kind of bonds: structure A (two Si$_{ad}$--Cl bonds, all Cl atoms
are bound with the surface), structure B (two Si$_{ad}$--Cl bonds,
only one Cl atom is bound with the surface), structure C (three
Si$_{ad}$--Cl bonds), and structure D (four Si$_{ad}$--Cl bonds). Structural parameters for all structures are
summarized in Table~\ref{table_bonds}.

\begin{table}[h]
\caption{Adsorption energies and lattice parameters for the structures shown in Fig.~\ref{fig1}: bond length between Si adatom (Si$_{ad}$) and the nearest Cl atom(s); bond length between Si surface atom (Si$_{s}$) and Cl atom of cluster formed by Si$_{ad}$;
bond length between Si$_{ad}$  and the nearest Si$_{s}$; and angle
between Si$_{ad}$ and the nearest Cl atoms of the cluster.}
 \label{table_bonds}
\begin{center}
\footnotesize{
   \begin{tabular}{|c| c | c | c |c|c|}   \hline
    Struct. & $E_{ads}$, eV& Si$_{ad}$--Cl, {\AA}  & Si$_{s}$--Cl, {\AA}  & Si$_{ad}$--Si$_{s}$, {\AA} & $\angle$(Cl-Si$_{ad}$-Cl)   \\ \hline \hline
    A & $-0.75$ & 2.36 & 2.25 & 3.94 & 96$^\circ$  \\ \hline
    A1 & $-0.54$ & 2.38 & 2.28 & 4.45 & 107$^\circ$  \\ \hline
    A2 & $-1.11$ & 2.10 & 3.75 & 5.10 & 102$^\circ$   \\ \hline
    B & $-1.51$ & Cl(1): 2.06 & 3.89 & 4.41 & 104$^\circ$    \\
       & & Cl(2): 2.18 & 2.57 &  &    \\ \hline
    B1 & $-1.30$ & Cl(1): 2.04 & 3.59 & 4.48 & 103$^\circ$    \\
       & & Cl(2): 2.21 & 2.48 &  &    \\ \hline
    B2 & $-1.18$ & Cl(1): 2.04 & 3.24 & 4.26 & 96$^\circ$    \\
       & & Cl(2): 2.42 & 2.26 &  &    \\ \hline
    C & $-1.77$ & Cl(1): 2.07 & 4.18 & 4.16 &99$^\circ$   \\
       & & Cl(2): 2.35 & 2.33 &  &    \\
       & & Cl(3): 2.55 & 2.20 &  &   \\ \hline
    C1 & $-1.72$ & Cl(1): 2.08 & 5.17 & 3.86 &90-95$^\circ$   \\
       & & Cl(2): 2.46 & 2.21 &  &   18, 19 \\
       & & Cl(3): 2.46 & 2.23 &  &   \\ \hline
    C2 & $-1.42$ & Cl(1): 2.05 & 4.06 & 3.86 &99-103$^\circ$   \\
       & & Cl(2): 2.33 & 2.33 &  &    \\
       & & Cl(3): 2.66 & 2.19 &  &    \\ \hline
    D & $-1.41$ & 2.70 & 2.14 & 3.18 &  89--91$^\circ$   \\ \hline
    D1 & $-0.26$ & 3.20 & 2.09 & 5.10 &  69--70$^\circ$ \\ \hline
    E & $-2.83$ & 2.09 & 3.68 & 2.50 & ---   \\ \hline
    F & $-2.89$ & 2.09 & 3.57 & 2.50 & ---   \\ \hline
    G & $-3.44$ & 2.05 & 3.91 & 2.44 & 106$^\circ$   \\ \hline
    H & $-3.24$ & 2.05  & 3.84 & 2.36 & 106$^\circ$  \\ \hline

\end{tabular}
}
\end{center}
\end{table}

In the structure A, a SiCl$_2^{\ast}$ cluster is formed (asterisk
denotes a cluster attached to the surface via Cl atom(s)). In the
cluster, the Si adatom forms two equivalent bonds with Cl atoms
belonging to the same silicon dimer. Structural parameters of the resulting SiCl$_2^{\ast}$
cluster appear to be equal to those calculated for the \ce{SiCl2} molecule in vacuum (Si--Cl bond length =
2.09\,{\AA}, $\angle$(Cl-Si$_{ad}$-Cl) = 102$^\circ$).

In the structure B, one chlorine atom of SiCl$_2^{\ast}$ cluster
forms a stronger bond with the surface, therefore the cluster
structure is slightly different from the free \ce{SiCl2} molecule.
SiCl$_2^{\ast}$ cluster can also be formed with Cl belonging to
different silicon dimers (structure B2).

SiCl$_3^{\ast}$ and SiCl$_4^{\ast}$ clusters formation is also
possible (structures C and D, correspondingly). The SiCl$_3^{\ast}$
cluster is the most stable one among other SiCl$_x^{\ast} (x=2, 3,
4)$ clusters attached to the surface via Cl atom(s).

\begin{figure}[h]
\begin{center}
    \includegraphics[width=\linewidth]{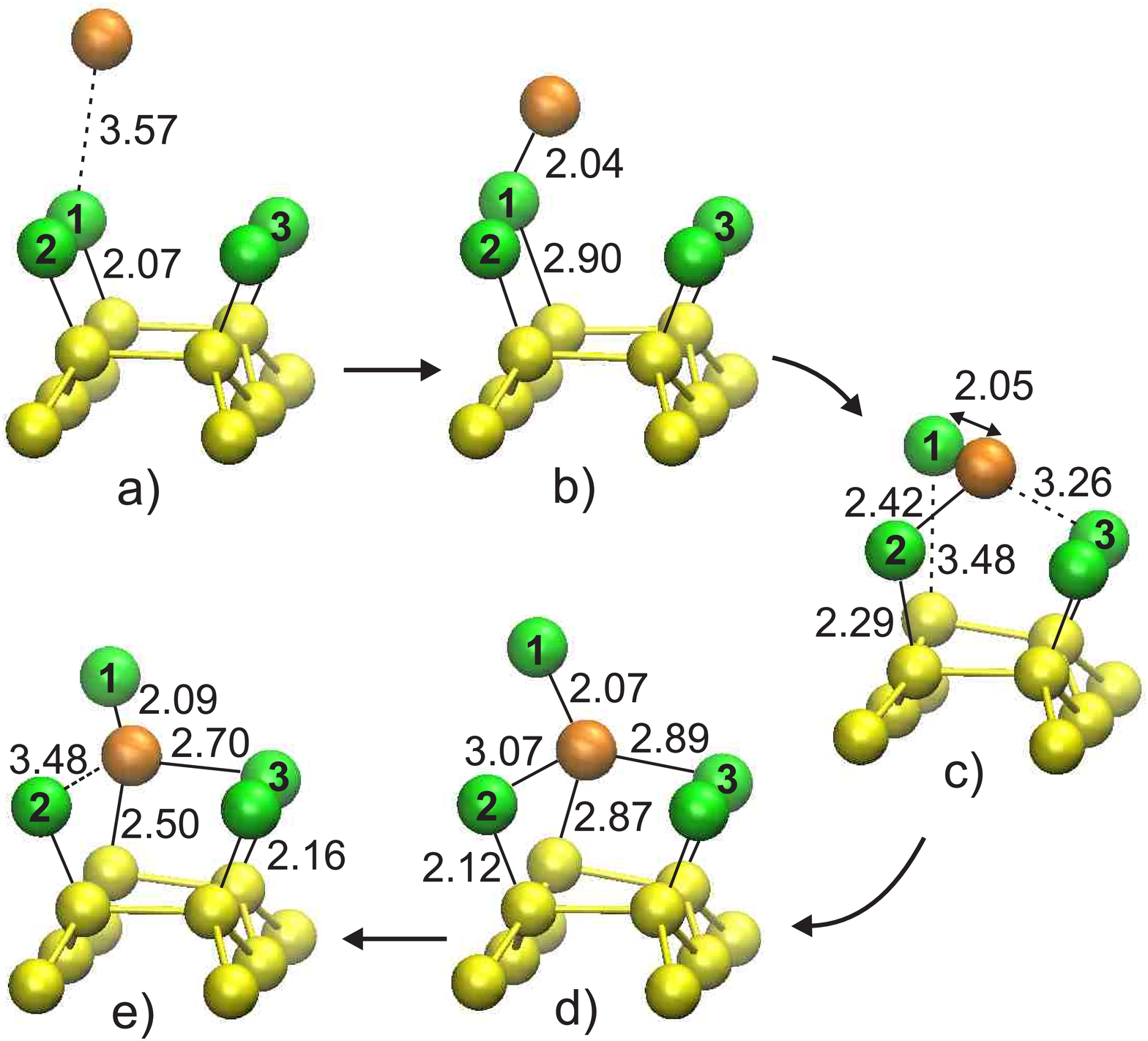}
    \caption{The process of geometry optimization for spontaneous substitutional adsorption. a) Initial structure; b-d) three structures through which the process of coordinates relaxation passes; e) optimized structure E.}
    \label{fig2}
\end{center}
\end{figure}

In the type II of adsorption structures, a SiCl$_x (x=1, 2)$ cluster
is attached to the surface via Si$_{ad}$--Si$_{s}$ bond(s) (such
clusters are denoted without asterisk). In the structure E
(Fig.~\ref{fig1}), Si adatom is bound with the silicon
lattice, and the chlorine segregates on top of the surface
structure. To obtain the structure E, the structure with a Si adatom at a distance of
3.5\,{\AA} above the chlorine monolayer was optimized using VASP code. The lateral position of Si$_{ad}$ was shifted from the top position of the Cl atom in the directions $\mathrm{[110]}$ (towards the dimer) and $\mathrm{[\bar{1}10]}$ by 0.50\,{\AA}. Figure~\ref{fig2} shows the initial structure (Fig.~\ref{fig2}a), the three structures through which the process of coordinates relaxation passes (Fig.~\ref{fig2}b-d), and the optimized structure E (Fig.~\ref{fig2}e). (From the set of structures, we have chosen the three structures in which the formation of bonds between the silicon atom and the neighboring chlorine atoms is most clearly seen.) At the beginning (Fig.~\ref{fig2}b), the Si adatom is attracted to
the nearest chlorine atom Cl(1), and the bond between chlorine
Cl(1) and the substrate Si atom becomes weaker.
Then, Si adatom exchanges places with chlorine Cl(1) and forms bonds
with the nearest chlorine atoms Cl(2) and Cl(3)
(Fig.~\ref{fig2}c,d). The process of exchanging chlorine and
silicon atoms occurs spontaneously. We believe that the energy required for Cl(1)--Si$_{s}$ bond breaking is compensated by the creation of bonds between Si$_{ad}$ and the neighboring chlorine atoms (Cl(2) and Cl(3)). In the final
position (Fig.~\ref{fig2}e), the bond length between Si adatom and
the nearest Si atom is slightly longer than that in the bulk
(2.50\,{\AA} vs 2.37\,{\AA}). It is worth mentioning, substitutional
adsorption leads to the most energetically favorable structure
($E_{ads} =-2.83$\,eV) among all other structures formed during
spontaneous adsorption of a silicon atom on Si(100)-2$\times$1-Cl.

Since such substitutional adsorption turned out to be very stable,
we have considered additional adsorption positions of a silicon atom
under the chlorine monolayer, which could be obtained by Si$_{ad}$
migration from its position in the structure E. The structure F is similar to E, but the position of Si$_{ad}$ is shifted closer to the row between the
dimers. The adsorption energies of the structures E and F containing
a SiCl cluster are approximately equal each other ($-2.83$\,eV and
$-2.89$\,eV, respectively).

In the most favorable configurations, Si$_{ad}$ forms \ce{SiCl2}
clusters in the inter-bridge and bridge dimer sites (structures G
and H in Fig.~\ref{fig1}). Atomic configuration G appears to be the
most stable structure of all considered in this paper. This result
strongly suggests that the formation of the Si$_{ad}$--Si$_{s}$
bonds makes a valuable contribution to the lowering of the adsorption
energy in comparison with the formation of Si$_{ad}$--Cl bonds only. Note that Si--Si surface dimer in model H is not tilted, while
in models G and E it is slightly tilted by 5$^\circ$ and 4$^\circ$,
respectively.

\begin{figure*}[t]
\begin{center}
    \includegraphics[width=0.8\linewidth]{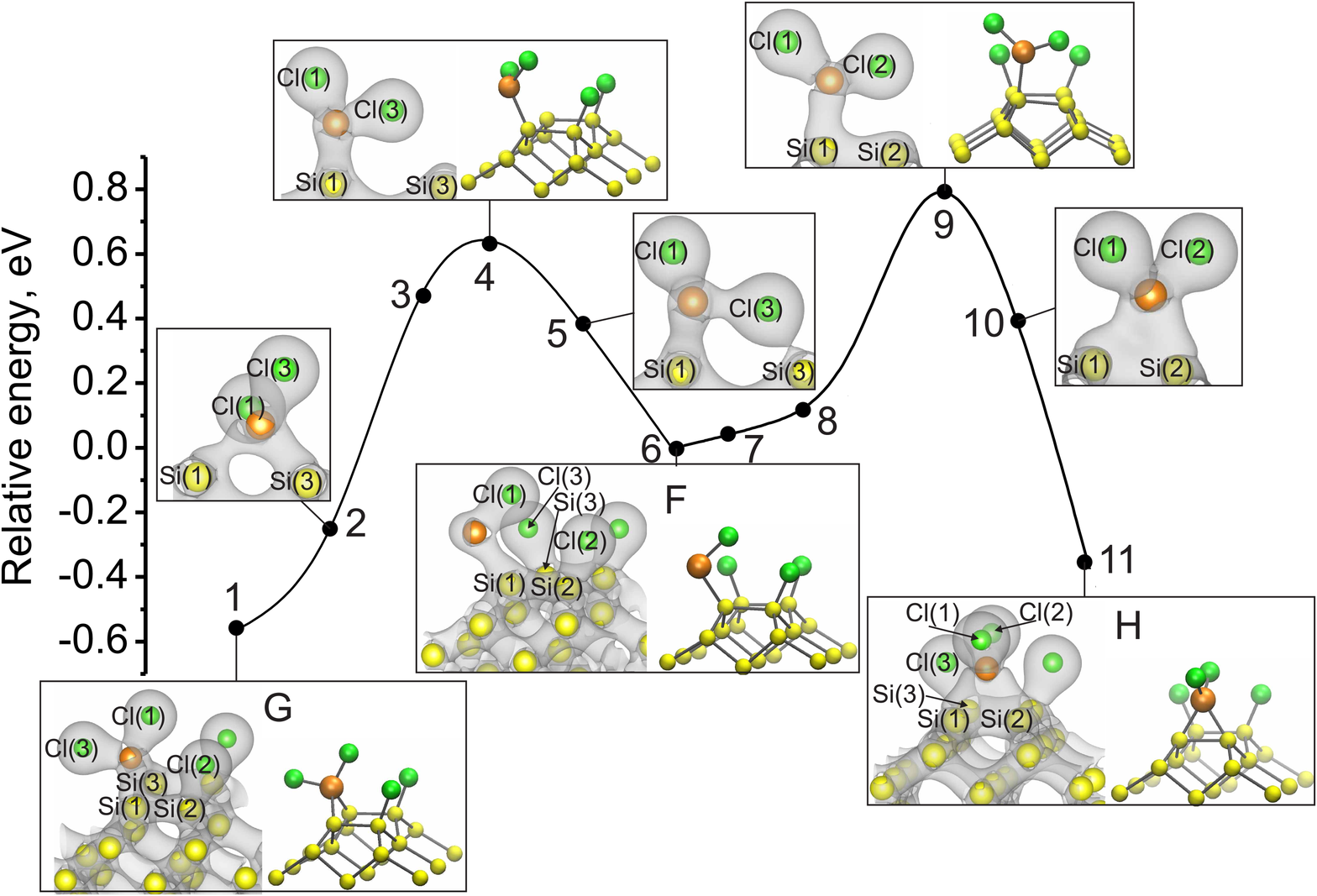}
    \caption{Reaction pathway for transitions from the structure F (image 6) to the most stable structures G (image 1) and
    H (image 11).Charge density distribution is shown for initial, final, and some intermediate states.}
    \label{fig3}
\end{center}
\end{figure*}

Transitions from the structure F to the most stable structures G and H are shown
in the energy diagram in Fig.~\ref{fig3}. The SiCl cluster in structure F
can attach the chlorine atom Cl(3) from the nearest dimer and
migrate to the most stable inter-bridge dimer site (structure G). To
break the bond between Cl(3) and the surface atom Si$_{s}(3)$, an
energy of 0.64\,eV is required. The charge density distribution in Fig.~\ref{fig3} clearly shows
Cl(3)--Si$_{s}(3)$ bond breaking (inserts 5$\rightarrow$4) and Si$_{ad}$--Si$_{s}(3)$ bond
formation (inserts 4$\rightarrow$2) . The reverse pathway (G$\rightarrow$F) requires a higher
activation energy, since a stronger bond between Si$_{ad}$ and
Si$_{s}$ should be broken. The transformation of the SiCl cluster in
the structure F to the \ce{SiCl2} cluster in the bridge dimer site
(structure H) is an alternative process. This process is not
energetically preferable, since the structure H is less favorable by
0.20\,eV and the activation barrier for the F$\rightarrow$H
transition is slightly higher (0.80\,eV) than that for the
F$\rightarrow$G transition (0.64\,eV).

\begin{figure*}[t]
\begin{center}
    \includegraphics[width=\linewidth]{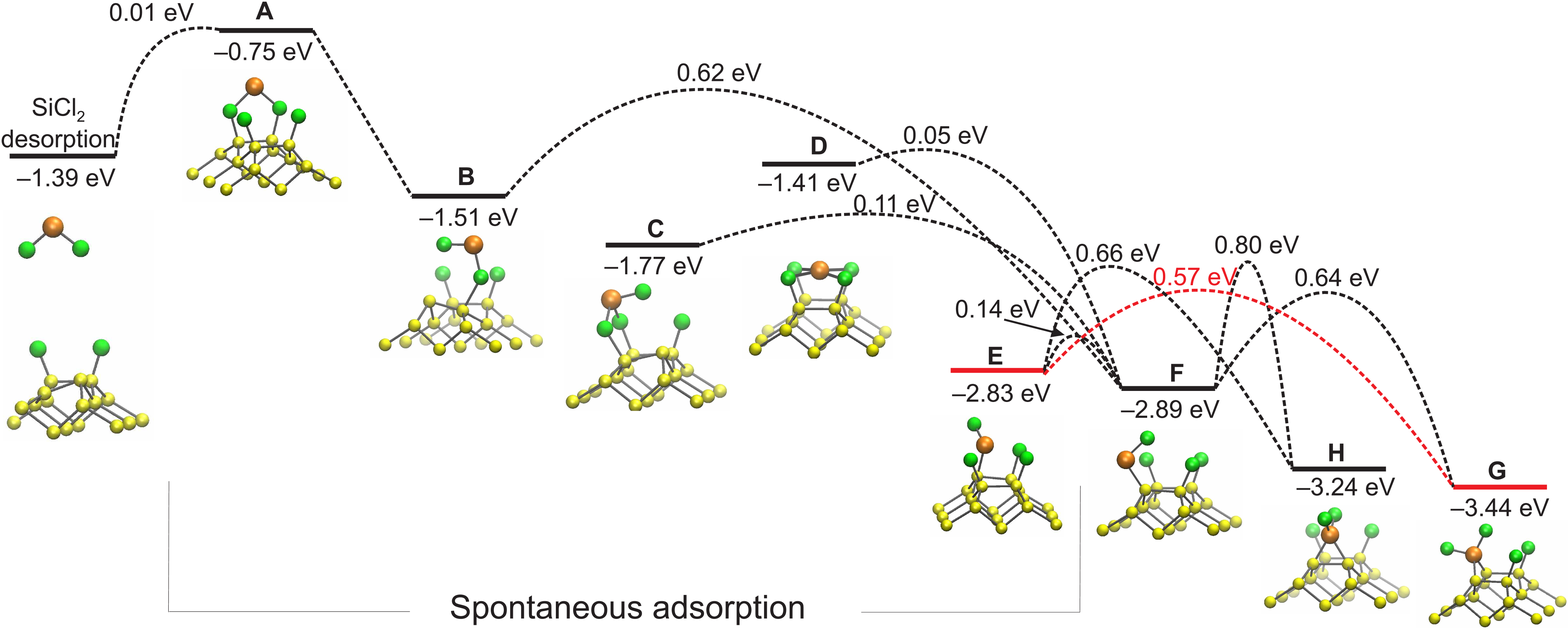}
    \caption{Transitions between adsorption structures A--G for Si$_{ad}$ on Si(100)-2$\times$1-Cl. The simplest way from spontaneous Si$_{ad}$ adsorption to the most stable adsorption site is marked by the red curve. Silicon atoms are yellow, chlorine atoms are green, and Si adatom is orange.}
    \label{fig4}
\end{center}
\end{figure*}

Transitions between different adsorption structures are shown in the energy diagram in
Fig.~\ref{fig4}. The SiCl$_2^{\ast}$ cluster in the structure A can
transfer to more stable structure B without an activation energy.
Further, the structure B can transform to the structure F with low
barrier (0.06\,eV). Structures C and D can also transform to the structure F with low
barrier (0.11 and 0.05\,eV, respectively). Pathway from the structures E to F
require activation energy of 0.14\,eV.  Transitions
E$\rightarrow$G (E$\rightarrow$H) requires approximately the same
activation energy as transitions F$\rightarrow$G (F$\rightarrow$H).
Thus, the simplest way from spontaneous Si$_{ad}$  adsorption to
the most stable adsorption site passes from the structure E to the structure
G and requires an energy of 0.57\,eV.

The SiCl$_2^{\ast}$ cluster formed in spontaneous Si$_{ad}$ adsorption can desorb as a \ce{SiCl2} molecule. Energies required for \ce{SiCl2} molecules desorption ($E_{des}$), as well as energy difference between final ($E_{f}$) and initial ($E_{i}$) states, are listed in Table~\ref{table_des}. The process of \ce{SiCl2} desorption from the structure A is exothermic and it can occur over a very low barrier ($E_{act} = 0.01$\,eV) (Fig.~\ref{fig4}). In the structure B, a \ce{SiCl2} molecule desorbs over a rather low barrier. Therefore, as soon as a SiCl$_2^{\ast}$ cluster forms on the surface, it can desorb as a \ce{SiCl2} molecule. The \ce{SiCl2} molecule desorption from the structures C and D requires higher activation barrier,  0.38\,eV and 1.38\,eV, correspondingly. Note that desorption of \ce{SiCl3} (\ce{SiCl4}) from the structure C (D) is less beneficial than the desorption of \ce{SiCl2} molecule.

\begin{table}[h]
\begin{center}
\caption{Energy difference between final ($E_{f} = -1.39$\,eV) and initial ($E_{i}$) states and energy barrier required for desorption. In the final state, a \ce{SiCl2} molecule is above the surface, and a double Cl vacancy (on one dimer) is formed on the surface. In the case of desorption process without activation energy, energy barrier coincides with the energy difference ($E_{f}-E_{i}$).}
    \label{table_des}    \small{
    \begin{tabular}{|c|c|c|c|}   \hline
    Model &  ($E_{f}-E_{i}$), eV & Energy barrier, eV  \\ \hline \hline
    A &  $-0.64$ & $0.01$  \\ \hline
    B &  $0.12$ & $0.12$   \\ \hline
    C &  $0.38$ &  $0.38$  \\ \hline
    D & $0.02$ & $1.38$  \\ \hline

\end{tabular}
}
\end{center}
\end{table}

\section{Discussion}

In this section, we turn to the comparison of Si adsorption on a
Cl-terminated Si(100)-2$\times$1 surface with a H-terminated one
\cite{1997Jeong}. In both cases, the Si adatom, spontaneously
substituting Z (Z=H, Cl) atom, can
capture one more Z atom and form a \ce{SiZ2} cluster in a bridge
dimer or an inter-bridge dimer site.

In the case of Si adsorption on hydrogenated surface at room
temperature, most of Si adatoms transfer from the substitutional
site to the bridge dimer site due to the low activation energy
(0.5\,eV) \cite{2005Kajiyama}. Further transition of the \ce{SiH2}
cluster into the most favorable position in the inter-bridge dimer
site requires an activation energy of 1.1\,eV \cite{2005Kajiyama},
which can hardly be overcome at room temperature. This scenario is
in agreement with the experiment \cite{2005Kajiyama}, where Si
adatoms deposited on a H-terminated Si(100)-2$\times$1 surface at
room temperature were at first found predominantly in the bridge
dimer sites, but after annealing to 250$^{\circ}$C --- in the
inter-bridge dimer sites.

According to our calculations of chlorinated surface, the minimum
energy path from the structure E to structure H requires an activation energy of
0.66\,eV, whereas the transition from the structure E to the most stable
structure (G) requires only 0.57\,eV. Thus, most Si adatoms
deposited on a Si(100)-2$\times$1-Cl surface at room temperature
should be adsorbed into the inter-bridge dimer site. According to
the silicon crystal structure, the right position for the growth of
a new layer is the bridge dimer site but not the inter-bridge dimer
site. However, at 250$^{\circ}$C, Si adatoms adsorbed on Si(100)-2$\times$1-H in the inter-bridge dimer sites do form a silicon
crystal structure (according to transmission electron microscopy
(TEM) experiments \cite{2016Deng,2018Wang}). To explain the
successful formation of a new crystalline layer from adatoms at the
inter-bridge dimer sites, it is necessary to study the process with
two or more Si adatoms.

We have found that the most stable structures are the same for Si
adsorbed on H- and Cl-terminated Si(100)-2$\times$1 and also the
activation barriers are comparable, so we can expect that the
epitaxy process will be similar. In the case of H-terminated
Si(100)-2$\times$1, it has been experimentally shown that silicon
epitaxy on hydrogen monolayer leads to hydrogen segregation to the
surface \cite{2016Deng}. Therefore it proves the possibility of Si
overgrowth on chlorine monolayer with Cl segregation to the surface
without Cl embedding in the epitaxial layers.

The difference between Si adsorption on Si(100)-2$\times$1-H and
Si(100)-2$\times$1-Cl is that in the later case, SiCl$_2^{\ast}$
clusters weakly bound to the Cl-terminated surface are formed. A
\ce{SiCl2} molecule can easily desorb from structures A and B
containing  SiCl$_2^{\ast}$ clusters, since the energy barriers
appear to be very low. As \ce{SiCl2} desorbs, chlorine is also partly removed from a Si(100)-2$\times$1-Cl
surface. In other words, the adsorbed silicon atoms etch the
chlorine layer but do not remove Si substrate atoms. The subsequent
Si overgrowth on the substrate area free of Cl should lead to a
more uniform growth of the epitaxial layers due to increased adatom
mobility on Cl-free surface (for example, Si adatom mobility is
lower on Si(100)-2$\times$1 surfaces terminated by hydrogen
\cite{1997Nara} and bromine \cite{2004Xu} than on a clean surface).
Note that the low-temperature removal of Cl will not lead to the
diffusion of dopants (for example, P atoms) on the surface.

\section{Conclusions}

The structures and energetics of Si adsorption on
Si(100)-2$\times$1-Cl have been studied with the density function
theory. The activation barriers for transitions between the most
stable states and desorption of  \ce{SiCl2} molecules from the
surface have also been calculated. When compared to the common
practice of using a hydrogen monolayer as a resist, a chlorine
monolayer presents not just similarities, but also potential
advantages due to low-temperature removal of chlorine during Si
epitaxy. Firstly, spontaneous substitutional adsorption takes place
both on H- \cite{2005Kajiyama} and Cl-terminated Si(100)-2$\times$1.
Secondly, the most stable sites for Si$_{ad}$ on both H-
\cite{2005Kajiyama} and Cl-terminated surfaces are inter-bridge and
bridge dimer sites, where Si$_{ad}$  is bound to
two Si and two H (or Cl) atoms. These results indicate that chlorine
segregates to the surface during Si deposition and is not
incorporated into the epitaxial layers. Thirdly, we have found out
that SiCl$_2^{\ast}$ clusters formed during silicon adsorption can
easily desorb from a Cl-terminated surface. Thus, depositing silicon
on a chlorine monolayer should produce silicon epitaxial layers of
quality at least not worse (may be even better) than on a hydrogen
monolayer.

\section{Acknowledgement}

The work was supported by Russian Science Foundation (grant
16-12-00050). We also thank the Joint Supercomputer Center of RAS
for providing the computing power.

\bibliographystyle{apsrev4-1}
\bibliography{T:/Manuscripts/Bibliography/Bibtex_Tania}

\begin{thebibliography}{32}%
\makeatletter
\providecommand \@ifxundefined [1]{%
 \@ifx{#1\undefined}
}%
\providecommand \@ifnum [1]{%
 \ifnum #1\expandafter \@firstoftwo
 \else \expandafter \@secondoftwo
 \fi
}%
\providecommand \@ifx [1]{%
 \ifx #1\expandafter \@firstoftwo
 \else \expandafter \@secondoftwo
 \fi
}%
\providecommand \natexlab [1]{#1}%
\providecommand \enquote  [1]{``#1''}%
\providecommand \bibnamefont  [1]{#1}%
\providecommand \bibfnamefont [1]{#1}%
\providecommand \citenamefont [1]{#1}%
\providecommand \href@noop [0]{\@secondoftwo}%
\providecommand \href [0]{\begingroup \@sanitize@url \@href}%
\providecommand \@href[1]{\@@startlink{#1}\@@href}%
\providecommand \@@href[1]{\endgroup#1\@@endlink}%
\providecommand \@sanitize@url [0]{\catcode `\\12\catcode `\$12\catcode
  `\&12\catcode `\#12\catcode `\^12\catcode `\_12\catcode `\%12\relax}%
\providecommand \@@startlink[1]{}%
\providecommand \@@endlink[0]{}%
\providecommand \url  [0]{\begingroup\@sanitize@url \@url }%
\providecommand \@url [1]{\endgroup\@href {#1}{\urlprefix }}%
\providecommand \urlprefix  [0]{URL }%
\providecommand \Eprint [0]{\href }%
\providecommand \doibase [0]{http://dx.doi.org/}%
\providecommand \selectlanguage [0]{\@gobble}%
\providecommand \bibinfo  [0]{\@secondoftwo}%
\providecommand \bibfield  [0]{\@secondoftwo}%
\providecommand \translation [1]{[#1]}%
\providecommand \BibitemOpen [0]{}%
\providecommand \bibitemStop [0]{}%
\providecommand \bibitemNoStop [0]{.\EOS\space}%
\providecommand \EOS [0]{\spacefactor3000\relax}%
\providecommand \BibitemShut  [1]{\csname bibitem#1\endcsname}%
\let\auto@bib@innerbib\@empty
\bibitem [{\citenamefont {Shen}\ \emph {et~al.}(2002)\citenamefont {Shen},
  \citenamefont {Ji}, \citenamefont {Zudov}, \citenamefont {Du}, \citenamefont
  {Kline},\ and\ \citenamefont {Tucker}}]{2002Shen}%
  \BibitemOpen
  \bibfield  {author} {\bibinfo {author} {\bibfnamefont {T.-C.}\ \bibnamefont
  {Shen}}, \bibinfo {author} {\bibfnamefont {J.-Y.}\ \bibnamefont {Ji}},
  \bibinfo {author} {\bibfnamefont {M.~A.}\ \bibnamefont {Zudov}}, \bibinfo
  {author} {\bibfnamefont {R.-R.}\ \bibnamefont {Du}}, \bibinfo {author}
  {\bibfnamefont {J.~S.}\ \bibnamefont {Kline}}, \ and\ \bibinfo {author}
  {\bibfnamefont {J.~R.}\ \bibnamefont {Tucker}},\ }\href {\doibase
  10.1063/1.1456949} {\bibfield  {journal} {\bibinfo  {journal} {Appl. Phys.
  Lett.}\ }\textbf {\bibinfo {volume} {80}},\ \bibinfo {pages} {1580} (\bibinfo
  {year} {2002})}\BibitemShut {NoStop}%
\bibitem [{\citenamefont {Keizer}\ \emph {et~al.}(2015)\citenamefont {Keizer},
  \citenamefont {Koelling}, \citenamefont {Koenraad},\ and\ \citenamefont
  {Simmons}}]{2015Keizer}%
  \BibitemOpen
  \bibfield  {author} {\bibinfo {author} {\bibfnamefont {J.}~\bibnamefont
  {Keizer}}, \bibinfo {author} {\bibfnamefont {S.}~\bibnamefont {Koelling}},
  \bibinfo {author} {\bibfnamefont {P.}~\bibnamefont {Koenraad}}, \ and\
  \bibinfo {author} {\bibfnamefont {M.}~\bibnamefont {Simmons}},\ }\href@noop
  {} {\bibfield  {journal} {\bibinfo  {journal} {ACS Nano}\ }\textbf {\bibinfo
  {volume} {9}},\ \bibinfo {pages} {2537} (\bibinfo {year} {2015})}\BibitemShut
  {NoStop}%
\bibitem [{\citenamefont {Weber}\ \emph {et~al.}(2012)\citenamefont {Weber},
  \citenamefont {Mahapatra}, \citenamefont {Ryu}, \citenamefont {Lee},
  \citenamefont {Fuhrer}, \citenamefont {Reusch}, \citenamefont {L.~Thompson},
  \citenamefont {C.~T.~Lee}, \citenamefont {Klimeck}, \citenamefont
  {C.~L.~Hollenberg},\ and\ \citenamefont {Simmons}}]{2012Weber}%
  \BibitemOpen
  \bibfield  {author} {\bibinfo {author} {\bibfnamefont {B.}~\bibnamefont
  {Weber}}, \bibinfo {author} {\bibfnamefont {S.}~\bibnamefont {Mahapatra}},
  \bibinfo {author} {\bibfnamefont {H.}~\bibnamefont {Ryu}}, \bibinfo {author}
  {\bibfnamefont {S.}~\bibnamefont {Lee}}, \bibinfo {author} {\bibfnamefont
  {A.}~\bibnamefont {Fuhrer}}, \bibinfo {author} {\bibfnamefont
  {T.}~\bibnamefont {Reusch}}, \bibinfo {author} {\bibfnamefont
  {D.}~\bibnamefont {L.~Thompson}}, \bibinfo {author} {\bibfnamefont
  {W.}~\bibnamefont {C.~T.~Lee}}, \bibinfo {author} {\bibfnamefont
  {G.}~\bibnamefont {Klimeck}}, \bibinfo {author} {\bibfnamefont
  {L.}~\bibnamefont {C.~L.~Hollenberg}}, \ and\ \bibinfo {author}
  {\bibfnamefont {M.}~\bibnamefont {Simmons}},\ }\href@noop {} {\bibfield
  {journal} {\bibinfo  {journal} {Science}\ }\textbf {\bibinfo {volume}
  {335}},\ \bibinfo {pages} {64} (\bibinfo {year} {2012})}\BibitemShut
  {NoStop}%
\bibitem [{\citenamefont {Weber}\ \emph {et~al.}(2014)\citenamefont {Weber},
  \citenamefont {Ryu}, \citenamefont {Tan}, \citenamefont {Klimeck},\ and\
  \citenamefont {Simmons}}]{2014Weber}%
  \BibitemOpen
  \bibfield  {author} {\bibinfo {author} {\bibfnamefont {B.}~\bibnamefont
  {Weber}}, \bibinfo {author} {\bibfnamefont {H.}~\bibnamefont {Ryu}}, \bibinfo
  {author} {\bibfnamefont {Y.-H.~M.}\ \bibnamefont {Tan}}, \bibinfo {author}
  {\bibfnamefont {G.}~\bibnamefont {Klimeck}}, \ and\ \bibinfo {author}
  {\bibfnamefont {M.~Y.}\ \bibnamefont {Simmons}},\ }\href {\doibase
  10.1103/PhysRevLett.113.246802} {\bibfield  {journal} {\bibinfo  {journal}
  {Phys. Rev. Lett.}\ }\textbf {\bibinfo {volume} {113}},\ \bibinfo {pages}
  {246802} (\bibinfo {year} {2014})}\BibitemShut {NoStop}%
\bibitem [{\citenamefont {Fuechsle}\ \emph {et~al.}(2012)\citenamefont
  {Fuechsle}, \citenamefont {Miwa}, \citenamefont {Mahapatra}, \citenamefont
  {Ryu}, \citenamefont {Lee}, \citenamefont {Warschkow}, \citenamefont
  {C~L~Hollenberg}, \citenamefont {Klimeck},\ and\ \citenamefont
  {Simmons}}]{2012Fuechsle}%
  \BibitemOpen
  \bibfield  {author} {\bibinfo {author} {\bibfnamefont {M.}~\bibnamefont
  {Fuechsle}}, \bibinfo {author} {\bibfnamefont {J.}~\bibnamefont {Miwa}},
  \bibinfo {author} {\bibfnamefont {S.}~\bibnamefont {Mahapatra}}, \bibinfo
  {author} {\bibfnamefont {H.}~\bibnamefont {Ryu}}, \bibinfo {author}
  {\bibfnamefont {S.}~\bibnamefont {Lee}}, \bibinfo {author} {\bibfnamefont
  {O.}~\bibnamefont {Warschkow}}, \bibinfo {author} {\bibfnamefont
  {L.}~\bibnamefont {C~L~Hollenberg}}, \bibinfo {author} {\bibfnamefont
  {G.}~\bibnamefont {Klimeck}}, \ and\ \bibinfo {author} {\bibfnamefont
  {M.}~\bibnamefont {Simmons}},\ }\href@noop {} {\bibfield  {journal} {\bibinfo
   {journal} {Nat. Nanotechnol.}\ }\textbf {\bibinfo {volume} {7}},\ \bibinfo
  {pages} {242} (\bibinfo {year} {2012})}\BibitemShut {NoStop}%
\bibitem [{\citenamefont {Oberbeck}\ \emph {et~al.}(2002)\citenamefont
  {Oberbeck}, \citenamefont {Curson}, \citenamefont {Simmons}, \citenamefont
  {Brenner}, \citenamefont {Hamilton}, \citenamefont {Schofield},\ and\
  \citenamefont {Clark}}]{2002Oberbeck}%
  \BibitemOpen
  \bibfield  {author} {\bibinfo {author} {\bibfnamefont {L.}~\bibnamefont
  {Oberbeck}}, \bibinfo {author} {\bibfnamefont {N.~J.}\ \bibnamefont
  {Curson}}, \bibinfo {author} {\bibfnamefont {M.~Y.}\ \bibnamefont {Simmons}},
  \bibinfo {author} {\bibfnamefont {R.}~\bibnamefont {Brenner}}, \bibinfo
  {author} {\bibfnamefont {A.~R.}\ \bibnamefont {Hamilton}}, \bibinfo {author}
  {\bibfnamefont {S.~R.}\ \bibnamefont {Schofield}}, \ and\ \bibinfo {author}
  {\bibfnamefont {R.~G.}\ \bibnamefont {Clark}},\ }\href {\doibase
  10.1063/1.1516859} {\bibfield  {journal} {\bibinfo  {journal} {Appl. Phys.
  Lett.}\ }\textbf {\bibinfo {volume} {81}},\ \bibinfo {pages} {3197} (\bibinfo
  {year} {2002})}\BibitemShut {NoStop}%
\bibitem [{\citenamefont {A.~Broome}\ \emph {et~al.}(2018)\citenamefont
  {A.~Broome}, \citenamefont {Gorman}, \citenamefont {House}, \citenamefont
  {Hile}, \citenamefont {Keizer}, \citenamefont {Keith}, \citenamefont
  {D.~Hill}, \citenamefont {Watson}, \citenamefont {Baker}, \citenamefont
  {C.~L.~Hollenberg},\ and\ \citenamefont {Simmons}}]{2018Broome}%
  \BibitemOpen
  \bibfield  {author} {\bibinfo {author} {\bibfnamefont {M.}~\bibnamefont
  {A.~Broome}}, \bibinfo {author} {\bibfnamefont {S.}~\bibnamefont {Gorman}},
  \bibinfo {author} {\bibfnamefont {M.}~\bibnamefont {House}}, \bibinfo
  {author} {\bibfnamefont {S.}~\bibnamefont {Hile}}, \bibinfo {author}
  {\bibfnamefont {J.}~\bibnamefont {Keizer}}, \bibinfo {author} {\bibfnamefont
  {D.}~\bibnamefont {Keith}}, \bibinfo {author} {\bibfnamefont
  {C.}~\bibnamefont {D.~Hill}}, \bibinfo {author} {\bibfnamefont
  {T.}~\bibnamefont {Watson}}, \bibinfo {author} {\bibfnamefont
  {W.}~\bibnamefont {Baker}}, \bibinfo {author} {\bibfnamefont
  {L.}~\bibnamefont {C.~L.~Hollenberg}}, \ and\ \bibinfo {author}
  {\bibfnamefont {M.}~\bibnamefont {Simmons}},\ }\href@noop {} {\bibfield
  {journal} {\bibinfo  {journal} {Nat. Commun.}\ }\textbf {\bibinfo {volume}
  {9}},\ \bibinfo {pages} {980} (\bibinfo {year} {2018})}\BibitemShut {NoStop}%
\bibitem [{\citenamefont {Deng}\ \emph {et~al.}(2016)\citenamefont {Deng},
  \citenamefont {Namboodiri}, \citenamefont {Li}, \citenamefont {Wang},
  \citenamefont {Stan}, \citenamefont {Myers}, \citenamefont {Cheng},
  \citenamefont {Li},\ and\ \citenamefont {Silver}}]{2016Deng}%
  \BibitemOpen
  \bibfield  {author} {\bibinfo {author} {\bibfnamefont {X.}~\bibnamefont
  {Deng}}, \bibinfo {author} {\bibfnamefont {P.}~\bibnamefont {Namboodiri}},
  \bibinfo {author} {\bibfnamefont {K.}~\bibnamefont {Li}}, \bibinfo {author}
  {\bibfnamefont {X.}~\bibnamefont {Wang}}, \bibinfo {author} {\bibfnamefont
  {G.}~\bibnamefont {Stan}}, \bibinfo {author} {\bibfnamefont {A.~F.}\
  \bibnamefont {Myers}}, \bibinfo {author} {\bibfnamefont {X.}~\bibnamefont
  {Cheng}}, \bibinfo {author} {\bibfnamefont {T.}~\bibnamefont {Li}}, \ and\
  \bibinfo {author} {\bibfnamefont {R.~M.}\ \bibnamefont {Silver}},\ }\href
  {\doibase https://doi.org/10.1016/j.apsusc.2016.03.212} {\bibfield  {journal}
  {\bibinfo  {journal} {Appl. Surf. Sci.}\ }\textbf {\bibinfo {volume} {378}},\
  \bibinfo {pages} {301} (\bibinfo {year} {2016})}\BibitemShut {NoStop}%
\bibitem [{\citenamefont {Hagmann}\ \emph {et~al.}(2018)\citenamefont
  {Hagmann}, \citenamefont {Wang}, \citenamefont {Namboodiri}, \citenamefont
  {Wyrick}, \citenamefont {Murray}, \citenamefont {Stewart}, \citenamefont
  {Silver},\ and\ \citenamefont {Richter}}]{2018Hagmann}%
  \BibitemOpen
  \bibfield  {author} {\bibinfo {author} {\bibfnamefont {J.~A.}\ \bibnamefont
  {Hagmann}}, \bibinfo {author} {\bibfnamefont {X.}~\bibnamefont {Wang}},
  \bibinfo {author} {\bibfnamefont {P.}~\bibnamefont {Namboodiri}}, \bibinfo
  {author} {\bibfnamefont {J.}~\bibnamefont {Wyrick}}, \bibinfo {author}
  {\bibfnamefont {R.}~\bibnamefont {Murray}}, \bibinfo {author} {\bibfnamefont
  {M.~D.}\ \bibnamefont {Stewart}}, \bibinfo {author} {\bibfnamefont {R.~M.}\
  \bibnamefont {Silver}}, \ and\ \bibinfo {author} {\bibfnamefont {C.~A.}\
  \bibnamefont {Richter}},\ }\href {\doibase 10.1063/1.4998712} {\bibfield
  {journal} {\bibinfo  {journal} {Appl. Phys. Lett.}\ }\textbf {\bibinfo
  {volume} {112}},\ \bibinfo {pages} {043102} (\bibinfo {year}
  {2018})}\BibitemShut {NoStop}%
\bibitem [{\citenamefont {Wang}\ \emph {et~al.}(2018)\citenamefont {Wang},
  \citenamefont {Hagmann}, \citenamefont {Namboodiri}, \citenamefont {Wyrick},
  \citenamefont {Li}, \citenamefont {Murray}, \citenamefont {Myers},
  \citenamefont {Misenkosen}, \citenamefont {Stewart}, \citenamefont
  {Richter},\ and\ \citenamefont {Silver}}]{2018Wang}%
  \BibitemOpen
  \bibfield  {author} {\bibinfo {author} {\bibfnamefont {X.}~\bibnamefont
  {Wang}}, \bibinfo {author} {\bibfnamefont {J.~A.}\ \bibnamefont {Hagmann}},
  \bibinfo {author} {\bibfnamefont {P.}~\bibnamefont {Namboodiri}}, \bibinfo
  {author} {\bibfnamefont {J.}~\bibnamefont {Wyrick}}, \bibinfo {author}
  {\bibfnamefont {K.}~\bibnamefont {Li}}, \bibinfo {author} {\bibfnamefont
  {R.~E.}\ \bibnamefont {Murray}}, \bibinfo {author} {\bibfnamefont
  {A.}~\bibnamefont {Myers}}, \bibinfo {author} {\bibfnamefont
  {F.}~\bibnamefont {Misenkosen}}, \bibinfo {author} {\bibfnamefont {M.~D.}\
  \bibnamefont {Stewart}}, \bibinfo {author} {\bibfnamefont {C.~A.}\
  \bibnamefont {Richter}}, \ and\ \bibinfo {author} {\bibfnamefont {R.~M.}\
  \bibnamefont {Silver}},\ }\href {\doibase 10.1039/C7NR07777G} {\bibfield
  {journal} {\bibinfo  {journal} {Nanoscale}\ }\textbf {\bibinfo {volume}
  {10}},\ \bibinfo {pages} {4488} (\bibinfo {year} {2018})}\BibitemShut
  {NoStop}%
\bibitem [{\citenamefont {Nara}\ \emph {et~al.}(1997)\citenamefont {Nara},
  \citenamefont {Sasaki},\ and\ \citenamefont {Ohno}}]{1997Nara}%
  \BibitemOpen
  \bibfield  {author} {\bibinfo {author} {\bibfnamefont {J.}~\bibnamefont
  {Nara}}, \bibinfo {author} {\bibfnamefont {T.}~\bibnamefont {Sasaki}}, \ and\
  \bibinfo {author} {\bibfnamefont {T.}~\bibnamefont {Ohno}},\ }\href {\doibase
  10.1103/PhysRevLett.79.4421} {\bibfield  {journal} {\bibinfo  {journal}
  {Phys. Rev. Lett.}\ }\textbf {\bibinfo {volume} {79}},\ \bibinfo {pages}
  {4421} (\bibinfo {year} {1997})}\BibitemShut {NoStop}%
\bibitem [{\citenamefont {Jeong}\ and\ \citenamefont
  {Oshiyama}(1997)}]{1997Jeong}%
  \BibitemOpen
  \bibfield  {author} {\bibinfo {author} {\bibfnamefont {S.}~\bibnamefont
  {Jeong}}\ and\ \bibinfo {author} {\bibfnamefont {A.}~\bibnamefont
  {Oshiyama}},\ }\href {\doibase 10.1103/PhysRevLett.79.4425} {\bibfield
  {journal} {\bibinfo  {journal} {Phys. Rev. Lett.}\ }\textbf {\bibinfo
  {volume} {79}},\ \bibinfo {pages} {4425} (\bibinfo {year}
  {1997})}\BibitemShut {NoStop}%
\bibitem [{\citenamefont {Jeong}\ and\ \citenamefont
  {Oshiyama}(1998)}]{1998Jeong}%
  \BibitemOpen
  \bibfield  {author} {\bibinfo {author} {\bibfnamefont {S.}~\bibnamefont
  {Jeong}}\ and\ \bibinfo {author} {\bibfnamefont {A.}~\bibnamefont
  {Oshiyama}},\ }\href {\doibase 10.1103/PhysRevB.58.12958} {\bibfield
  {journal} {\bibinfo  {journal} {Phys. Rev. B}\ }\textbf {\bibinfo {volume}
  {58}},\ \bibinfo {pages} {12958} (\bibinfo {year} {1998})}\BibitemShut
  {NoStop}%
\bibitem [{\citenamefont {Kajiyama}\ \emph {et~al.}(2005)\citenamefont
  {Kajiyama}, \citenamefont {Suwa}, \citenamefont {Heike}, \citenamefont
  {Fujimori}, \citenamefont {Nara}, \citenamefont {Ohno}, \citenamefont
  {Matsuura}, \citenamefont {Hitosugi},\ and\ \citenamefont
  {Hashizume}}]{2005Kajiyama}%
  \BibitemOpen
  \bibfield  {author} {\bibinfo {author} {\bibfnamefont {H.}~\bibnamefont
  {Kajiyama}}, \bibinfo {author} {\bibfnamefont {Y.}~\bibnamefont {Suwa}},
  \bibinfo {author} {\bibfnamefont {S.}~\bibnamefont {Heike}}, \bibinfo
  {author} {\bibfnamefont {M.}~\bibnamefont {Fujimori}}, \bibinfo {author}
  {\bibfnamefont {J.}~\bibnamefont {Nara}}, \bibinfo {author} {\bibfnamefont
  {T.}~\bibnamefont {Ohno}}, \bibinfo {author} {\bibfnamefont {S.}~\bibnamefont
  {Matsuura}}, \bibinfo {author} {\bibfnamefont {T.}~\bibnamefont {Hitosugi}},
  \ and\ \bibinfo {author} {\bibfnamefont {T.}~\bibnamefont {Hashizume}},\
  }\href {\doibase 10.1143/JPSJ.74.389} {\bibfield  {journal} {\bibinfo
  {journal} {J. Phys. Soc. Jpn.}\ }\textbf {\bibinfo {volume} {74}},\ \bibinfo
  {pages} {389} (\bibinfo {year} {2005})}\BibitemShut {NoStop}%
\bibitem [{\citenamefont {Ji}\ and\ \citenamefont {Shen}(2004)}]{2004Ji}%
  \BibitemOpen
  \bibfield  {author} {\bibinfo {author} {\bibfnamefont {J.-Y.}\ \bibnamefont
  {Ji}}\ and\ \bibinfo {author} {\bibfnamefont {T.-C.}\ \bibnamefont {Shen}},\
  }\href {\doibase 10.1103/PhysRevB.70.115309} {\bibfield  {journal} {\bibinfo
  {journal} {Phys. Rev. B}\ }\textbf {\bibinfo {volume} {70}},\ \bibinfo
  {pages} {115309} (\bibinfo {year} {2004})}\BibitemShut {NoStop}%
\bibitem [{\citenamefont {Moon}\ \emph {et~al.}(2007)\citenamefont {Moon},
  \citenamefont {Jeon}, \citenamefont {Hwang}, \citenamefont {Hwang},
  \citenamefont {Song}, \citenamefont {Shin}, \citenamefont {Chung},\ and\
  \citenamefont {Park}}]{2007Moon}%
  \BibitemOpen
  \bibfield  {author} {\bibinfo {author} {\bibfnamefont {S.}~\bibnamefont
  {Moon}}, \bibinfo {author} {\bibfnamefont {C.}~\bibnamefont {Jeon}}, \bibinfo
  {author} {\bibfnamefont {H.}~\bibnamefont {Hwang}}, \bibinfo {author}
  {\bibfnamefont {C.}~\bibnamefont {Hwang}}, \bibinfo {author} {\bibfnamefont
  {H.}~\bibnamefont {Song}}, \bibinfo {author} {\bibfnamefont {H.}~\bibnamefont
  {Shin}}, \bibinfo {author} {\bibfnamefont {S.}~\bibnamefont {Chung}}, \ and\
  \bibinfo {author} {\bibfnamefont {C.}~\bibnamefont {Park}},\ }\href {\doibase
  10.1002/adma.200602166} {\bibfield  {journal} {\bibinfo  {journal} {Adv.
  Mater.}\ }\textbf {\bibinfo {volume} {19}},\ \bibinfo {pages} {1321}
  (\bibinfo {year} {2007})}\BibitemShut {NoStop}%
\bibitem [{\citenamefont {Jeon}\ \emph {et~al.}(2011)\citenamefont {Jeon},
  \citenamefont {Hwang}, \citenamefont {Shin}, \citenamefont {Park},\ and\
  \citenamefont {Hwang}}]{2011Jeon}%
  \BibitemOpen
  \bibfield  {author} {\bibinfo {author} {\bibfnamefont {C.}~\bibnamefont
  {Jeon}}, \bibinfo {author} {\bibfnamefont {H.-N.}\ \bibnamefont {Hwang}},
  \bibinfo {author} {\bibfnamefont {H.-J.}\ \bibnamefont {Shin}}, \bibinfo
  {author} {\bibfnamefont {C.-Y.}\ \bibnamefont {Park}}, \ and\ \bibinfo
  {author} {\bibfnamefont {C.-C.}\ \bibnamefont {Hwang}},\ }\href {\doibase
  https://doi.org/10.1016/j.apsusc.2011.04.043} {\bibfield  {journal} {\bibinfo
   {journal} {Appl. Surf. Sci.}\ }\textbf {\bibinfo {volume} {257}},\ \bibinfo
  {pages} {8794 } (\bibinfo {year} {2011})}\BibitemShut {NoStop}%
\bibitem [{\citenamefont {de~Wijs}\ \emph {et~al.}(1998)\citenamefont
  {de~Wijs}, \citenamefont {De~Vita},\ and\ \citenamefont
  {Selloni}}]{1998deWijs}%
  \BibitemOpen
  \bibfield  {author} {\bibinfo {author} {\bibfnamefont {G.~A.}\ \bibnamefont
  {de~Wijs}}, \bibinfo {author} {\bibfnamefont {A.}~\bibnamefont {De~Vita}}, \
  and\ \bibinfo {author} {\bibfnamefont {A.}~\bibnamefont {Selloni}},\ }\href
  {\doibase 10.1103/PhysRevB.57.10021} {\bibfield  {journal} {\bibinfo
  {journal} {Phys. Rev. B}\ }\textbf {\bibinfo {volume} {57}},\ \bibinfo
  {pages} {10021} (\bibinfo {year} {1998})}\BibitemShut {NoStop}%
\bibitem [{\citenamefont {Aldao}\ \emph {et~al.}(2009)\citenamefont {Aldao},
  \citenamefont {Agrawal}, \citenamefont {Butera},\ and\ \citenamefont
  {Weaver}}]{2009Aldao}%
  \BibitemOpen
  \bibfield  {author} {\bibinfo {author} {\bibfnamefont {C.~M.}\ \bibnamefont
  {Aldao}}, \bibinfo {author} {\bibfnamefont {A.}~\bibnamefont {Agrawal}},
  \bibinfo {author} {\bibfnamefont {R.~E.}\ \bibnamefont {Butera}}, \ and\
  \bibinfo {author} {\bibfnamefont {J.~H.}\ \bibnamefont {Weaver}},\ }\href
  {\doibase 10.1103/PhysRevB.79.125303} {\bibfield  {journal} {\bibinfo
  {journal} {Phys. Rev. B}\ }\textbf {\bibinfo {volume} {79}},\ \bibinfo
  {pages} {125303} (\bibinfo {year} {2009})}\BibitemShut {NoStop}%
\bibitem [{\citenamefont {Pavlova}\ \emph {et~al.}(2018)\citenamefont
  {Pavlova}, \citenamefont {Zhidomirov},\ and\ \citenamefont
  {Eltsov}}]{2018Pavlova}%
  \BibitemOpen
  \bibfield  {author} {\bibinfo {author} {\bibfnamefont {T.~V.}\ \bibnamefont
  {Pavlova}}, \bibinfo {author} {\bibfnamefont {G.~M.}\ \bibnamefont
  {Zhidomirov}}, \ and\ \bibinfo {author} {\bibfnamefont {K.~N.}\ \bibnamefont
  {Eltsov}},\ }\href {\doibase 10.1021/acs.jpcc.7b11519} {\bibfield  {journal}
  {\bibinfo  {journal} {J. Phys. Chem. C}\ }\textbf {\bibinfo {volume} {122}},\
  \bibinfo {pages} {1741} (\bibinfo {year} {2018})}\BibitemShut {NoStop}%
\bibitem [{\citenamefont {Hall}\ \emph {et~al.}(2001)\citenamefont {Hall},
  \citenamefont {Mui},\ and\ \citenamefont {Musgrave}}]{2001Hall}%
  \BibitemOpen
  \bibfield  {author} {\bibinfo {author} {\bibfnamefont {M.~A.}\ \bibnamefont
  {Hall}}, \bibinfo {author} {\bibfnamefont {C.}~\bibnamefont {Mui}}, \ and\
  \bibinfo {author} {\bibfnamefont {C.~B.}\ \bibnamefont {Musgrave}},\ }\href
  {\doibase 10.1021/jp0118874} {\bibfield  {journal} {\bibinfo  {journal} {J.
  Phys. Chem. B}\ }\textbf {\bibinfo {volume} {105}},\ \bibinfo {pages} {12068}
  (\bibinfo {year} {2001})}\BibitemShut {NoStop}%
\bibitem [{\citenamefont {Kunioshi}\ \emph {et~al.}(2018)\citenamefont
  {Kunioshi}, \citenamefont {Fujimura}, \citenamefont {Fuwa},\ and\
  \citenamefont {Yamaguchi}}]{2018Kunioshi}%
  \BibitemOpen
  \bibfield  {author} {\bibinfo {author} {\bibfnamefont {N.}~\bibnamefont
  {Kunioshi}}, \bibinfo {author} {\bibfnamefont {Y.}~\bibnamefont {Fujimura}},
  \bibinfo {author} {\bibfnamefont {A.}~\bibnamefont {Fuwa}}, \ and\ \bibinfo
  {author} {\bibfnamefont {K.}~\bibnamefont {Yamaguchi}},\ }\href {\doibase
  https://doi.org/10.1016/j.commatsci.2018.08.037} {\bibfield  {journal}
  {\bibinfo  {journal} {Comp. Mater. Sci.}\ }\textbf {\bibinfo {volume}
  {155}},\ \bibinfo {pages} {28} (\bibinfo {year} {2018})}\BibitemShut
  {NoStop}%
\bibitem [{\citenamefont {Yadav}\ and\ \citenamefont
  {Singh}(2019)}]{2019Yadav}%
  \BibitemOpen
  \bibfield  {author} {\bibinfo {author} {\bibfnamefont {S.}~\bibnamefont
  {Yadav}}\ and\ \bibinfo {author} {\bibfnamefont {C.~V.}\ \bibnamefont
  {Singh}},\ }\href {\doibase https://doi.org/10.1016/j.apsusc.2018.12.253}
  {\bibfield  {journal} {\bibinfo  {journal} {Appl. Surf. Sci.}\ }\textbf
  {\bibinfo {volume} {475}},\ \bibinfo {pages} {124} (\bibinfo {year}
  {2019})}\BibitemShut {NoStop}%
\bibitem [{\citenamefont {Gao}\ and\ \citenamefont
  {Teplyakov}(2014)}]{2014Gao}%
  \BibitemOpen
  \bibfield  {author} {\bibinfo {author} {\bibfnamefont {F.}~\bibnamefont
  {Gao}}\ and\ \bibinfo {author} {\bibfnamefont {A.~V.}\ \bibnamefont
  {Teplyakov}},\ }\href {\doibase 10.1021/jp5095307} {\bibfield  {journal}
  {\bibinfo  {journal} {J. Phys. Chem. C}\ }\textbf {\bibinfo {volume} {118}},\
  \bibinfo {pages} {27998} (\bibinfo {year} {2014})}\BibitemShut {NoStop}%
\bibitem [{\citenamefont {Gao}\ and\ \citenamefont
  {Teplyakov}(2016)}]{2016Gao}%
  \BibitemOpen
  \bibfield  {author} {\bibinfo {author} {\bibfnamefont {F.}~\bibnamefont
  {Gao}}\ and\ \bibinfo {author} {\bibfnamefont {A.~V.}\ \bibnamefont
  {Teplyakov}},\ }\href {\doibase 10.1021/acs.jpcc.5b12424} {\bibfield
  {journal} {\bibinfo  {journal} {J. Phys. Chem. C}\ }\textbf {\bibinfo
  {volume} {120}},\ \bibinfo {pages} {5539} (\bibinfo {year}
  {2016})}\BibitemShut {NoStop}%
\bibitem [{\citenamefont {Lange}\ and\ \citenamefont
  {Schmidt}(2008)}]{2008Lange}%
  \BibitemOpen
  \bibfield  {author} {\bibinfo {author} {\bibfnamefont {B.}~\bibnamefont
  {Lange}}\ and\ \bibinfo {author} {\bibfnamefont {W.}~\bibnamefont
  {Schmidt}},\ }\href@noop {} {\bibfield  {journal} {\bibinfo  {journal} {Surf.
  Sci.}\ }\textbf {\bibinfo {volume} {602}},\ \bibinfo {pages} {1207} (\bibinfo
  {year} {2008})}\BibitemShut {NoStop}%
\bibitem [{\citenamefont {Soria}\ \emph {et~al.}(2013)\citenamefont {Soria},
  \citenamefont {Patrito},\ and\ \citenamefont {Paredes-Olivera}}]{2013Soria}%
  \BibitemOpen
  \bibfield  {author} {\bibinfo {author} {\bibfnamefont {F.~A.}\ \bibnamefont
  {Soria}}, \bibinfo {author} {\bibfnamefont {E.~M.}\ \bibnamefont {Patrito}},
  \ and\ \bibinfo {author} {\bibfnamefont {P.}~\bibnamefont
  {Paredes-Olivera}},\ }\href {\doibase 10.1021/jp4014042} {\bibfield
  {journal} {\bibinfo  {journal} {J. Phys. Chem. C}\ }\textbf {\bibinfo
  {volume} {117}},\ \bibinfo {pages} {18021} (\bibinfo {year}
  {2013})}\BibitemShut {NoStop}%
\bibitem [{\citenamefont {Kresse}\ and\ \citenamefont
  {Hafner}(1993)}]{1993Kresse}%
  \BibitemOpen
  \bibfield  {author} {\bibinfo {author} {\bibfnamefont {G.}~\bibnamefont
  {Kresse}}\ and\ \bibinfo {author} {\bibfnamefont {J.}~\bibnamefont
  {Hafner}},\ }\href {\doibase 10.1103/PhysRevB.47.558} {\bibfield  {journal}
  {\bibinfo  {journal} {Phys. Rev. B}\ }\textbf {\bibinfo {volume} {47}},\
  \bibinfo {pages} {558} (\bibinfo {year} {1993})}\BibitemShut {NoStop}%
\bibitem [{\citenamefont {Kresse}\ and\ \citenamefont
  {Furthm\"uller}(1996)}]{1996Kresse}%
  \BibitemOpen
  \bibfield  {author} {\bibinfo {author} {\bibfnamefont {G.}~\bibnamefont
  {Kresse}}\ and\ \bibinfo {author} {\bibfnamefont {J.}~\bibnamefont
  {Furthm\"uller}},\ }\href {\doibase 10.1103/PhysRevB.54.11169} {\bibfield
  {journal} {\bibinfo  {journal} {Phys. Rev. B}\ }\textbf {\bibinfo {volume}
  {54}},\ \bibinfo {pages} {11169} (\bibinfo {year} {1996})}\BibitemShut
  {NoStop}%
\bibitem [{\citenamefont {Perdew}\ \emph {et~al.}(1996)\citenamefont {Perdew},
  \citenamefont {Burke},\ and\ \citenamefont {Ernzerhof}}]{1996Perdew}%
  \BibitemOpen
  \bibfield  {author} {\bibinfo {author} {\bibfnamefont {J.~P.}\ \bibnamefont
  {Perdew}}, \bibinfo {author} {\bibfnamefont {K.}~\bibnamefont {Burke}}, \
  and\ \bibinfo {author} {\bibfnamefont {M.}~\bibnamefont {Ernzerhof}},\ }\href
  {\doibase 10.1103/PhysRevLett.77.3865} {\bibfield  {journal} {\bibinfo
  {journal} {Phys. Rev. Lett.}\ }\textbf {\bibinfo {volume} {77}},\ \bibinfo
  {pages} {3865} (\bibinfo {year} {1996})}\BibitemShut {NoStop}%
\bibitem [{\citenamefont {Henkelman}\ \emph {et~al.}(2000)\citenamefont
  {Henkelman}, \citenamefont {Uberuaga},\ and\ \citenamefont
  {J{\'o}nsson}}]{2000CNEB}%
  \BibitemOpen
  \bibfield  {author} {\bibinfo {author} {\bibfnamefont {G.}~\bibnamefont
  {Henkelman}}, \bibinfo {author} {\bibfnamefont {B.~P.}\ \bibnamefont
  {Uberuaga}}, \ and\ \bibinfo {author} {\bibfnamefont {H.}~\bibnamefont
  {J{\'o}nsson}},\ }\href {\doibase 10.1063/1.1329672} {\bibfield  {journal}
  {\bibinfo  {journal} {J. Chem. Phys.}\ }\textbf {\bibinfo {volume} {113}},\
  \bibinfo {pages} {9901} (\bibinfo {year} {2000})}\BibitemShut {NoStop}%
\bibitem [{\citenamefont {Xu}\ and\ \citenamefont {Weaver}(2004)}]{2004Xu}%
  \BibitemOpen
  \bibfield  {author} {\bibinfo {author} {\bibfnamefont {G.~J.}\ \bibnamefont
  {Xu}}\ and\ \bibinfo {author} {\bibfnamefont {J.~H.}\ \bibnamefont
  {Weaver}},\ }\href {\doibase 10.1103/PhysRevB.70.165321} {\bibfield
  {journal} {\bibinfo  {journal} {Phys. Rev. B}\ }\textbf {\bibinfo {volume}
  {70}},\ \bibinfo {pages} {165321} (\bibinfo {year} {2004})}\BibitemShut
  {NoStop}%
\end{thebibliography}%

\end{document}